\documentstyle[psfig,namedreferences,amssymb]{spacekap}

\runningtitle{Measurement of sky clarity using FIR radiometers}

\begin{opening}

\title{Measurement of sky clarity using FIR radiometers as an adjunct to
atmospheric \v{C}erenkov radiation measurements}

\author{D. J. \surname{Buckley}$^1$}
\author{M. C. \surname{Dorrington}$^1$}
\author{P. J. \surname{Edwards}$^1$} 
\author{T. J. L. \surname{McComb}$^{1,2}$}
\author{S. P. \surname{Tummey}$^1$} 
\author{K. E. \surname{Turver}$^{1,2}$}

\institute{$^{1}$University of Durham Industrial Research Laboratories,
Mountjoy Research Centre, University of Durham, Stockton Road, Durham
DH1 3UP, UK}

\institute{$^{2}$Department of Physics, Rochester Building, Science
Laboratories, University of Durham, Durham DH1 3LE, U.K.}

\date{Submitted to Experimental Astronomy, 30th October 1998}

\end{opening}

\begin{ao}

Prof. K. E. Turver, 
Department of Physics, 
Rochester Building, 
Science Laboratories, 
University of Durham, 
Durham DH1 3LE, 
UK.
e-mail k.e.turver@dur.ac.uk

\end{ao}

\begin{document}

\begin{abstract} A technique for detecting the presence of cloud in the
field of view of an atmospheric \v{C}erenkov telescope using a far infra
red radiometer is described. Models for the radiative emission from
clear and cloudy skies are tested and found to represent the
measurements.


\end{abstract}

\section{Introduction}

Measurement of atmospheric \v{C}erenkov radiation is an important
technique finding increasing application in high energy astrophysics
\cite{Fegan1997}. Observations are customarily made under conditions of
cloudless and moonless dark skies. A long standing problem has been the
need to monitor the clarity of the atmosphere --- the radiating medium
of the \v{C}erenkov light production and detection system. A number of
possibilities exist; recent suggestions, triggered by plans for future
large air shower recorders, employ active systems involving laser
probing of the atmosphere and measurements of the scattered light
\cite{Abuzayyad1997} or the monitoring of a complete hemisphere using a
matrix of fixed far infra red (FIR) radiometers \cite{Bird1997}.
Alternatively, monitoring of the brightness of a star in the field of a
gamma ray telescope is a passive method already employed for ground
based gamma ray astronomy \cite{Armstrong1998}. However, this method has
not found general application because measurements of faint stars using
inexpensive CCD detectors tend to be noisy giving an imprecise estimate
of atmospheric clarity. Often the performance of the gamma ray telescope
when measuring the \v{C}erenkov radiation produced by the cosmic ray
background is used as a monitor of the transparency of the atmosphere.
This has the disadvantage of depending on the performance of the
\v{C}erenkov detector itself.

We report here a technique involving measuring the radiative temperature
of the night sky and sensing the presence of cloud and other obscurants
against the cold clear sky. The method is quick, sensitive and
independent of the performance of the atmospheric \v{C}erenkov detector.

We present a range of observational radiometric data for the temperature
of clear and cloudy skies and describe these data using two simple
models for the temperature of clear sky and of clouds at various
heights.

\section{Experimental Equipment}

Measurements of the sky temperature have been made with FIR radiometers
(Heimann model KT 17 and KT 19) sensitive in the 8 -- 14 $\mu$m waveband
which coincides with an atmospheric window. The radiometers have a
temperature range $-75^\circ$C -- $+100^\circ$C, a temperatue resolution
of $\pm 0.2^\circ$C, and an aperture of $2^\circ$ defined by a germanium
lens. The radiometers were calibrated including allowance for the
effects of the instrument temperature on the performance of the
radiometer ($\sim 1^\circ$C change in indicated temperature for a
$10^\circ$C change in the temperature of the instument). 

Measurements were made with radiometers mounted on the Mark 6 gamma ray
telescope \cite{Armstrong1998} in Narrabri, NSW, Australia and on
steerable alt-azimuth platforms and on fixed mounts in Durham, England.
Data from the measurements made at various angles to the zenith,
throughout the day and night during different seasons over a period of
years are available. Cloud base measurements were made in 1998 using a
LIDAR ceilometer for a subset of the data. Full meteorological data are
available for all observations.

\section{Results}

\subsection{Correlation between telescope background count rate and sky
radiation temperature}

\begin{figure}

\centerline{\psfig{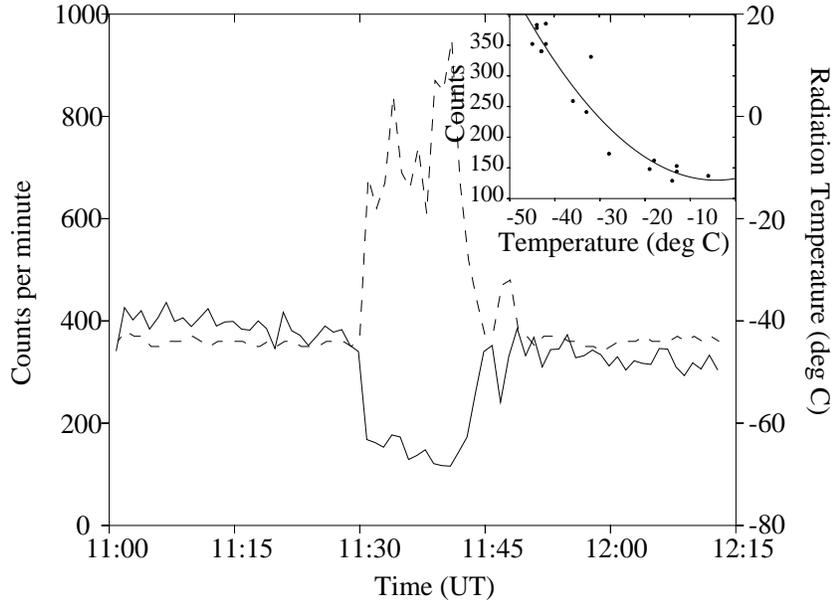}}

\caption{The correlation between the background counting rate of the
Mark 6 gamma ray telescope (solid line) and the radiative temperature of the
sky (broken line).}\label{countrate}

\end{figure}

The performance of a gamma ray telescope and the stability of the
atmosphere has often been monitored on the basis of the background
cosmic ray count rate of the telescope. This method has a number of
disadvantages. It depends upon the integrity of the telescope
performance and it is governed by the statistics of counting (at rates
with current telescopes up to 1000 cpm). We show in
Figure~\ref{countrate} the correlation between the background count rate
of the telescope and the radiative temperature of the sky during an
observation using the Mark 6 telescope at Narrabri. The sky temperature
is seen to be very sensitive to the presence of the warmer clouds and
correlates with the background count rate of the telescope. The presence
of clouds caused a decrease in telescope count rate and an increase in
sky temperature. It thus presents an instantaneous, sensitive and
independent monitor of the clarity of the sky in the beam of the gamma
ray telescope.

\subsection{Zenith angle dependence of temperature of a clear sky}

\begin{figure}

\centerline{\psfig{file=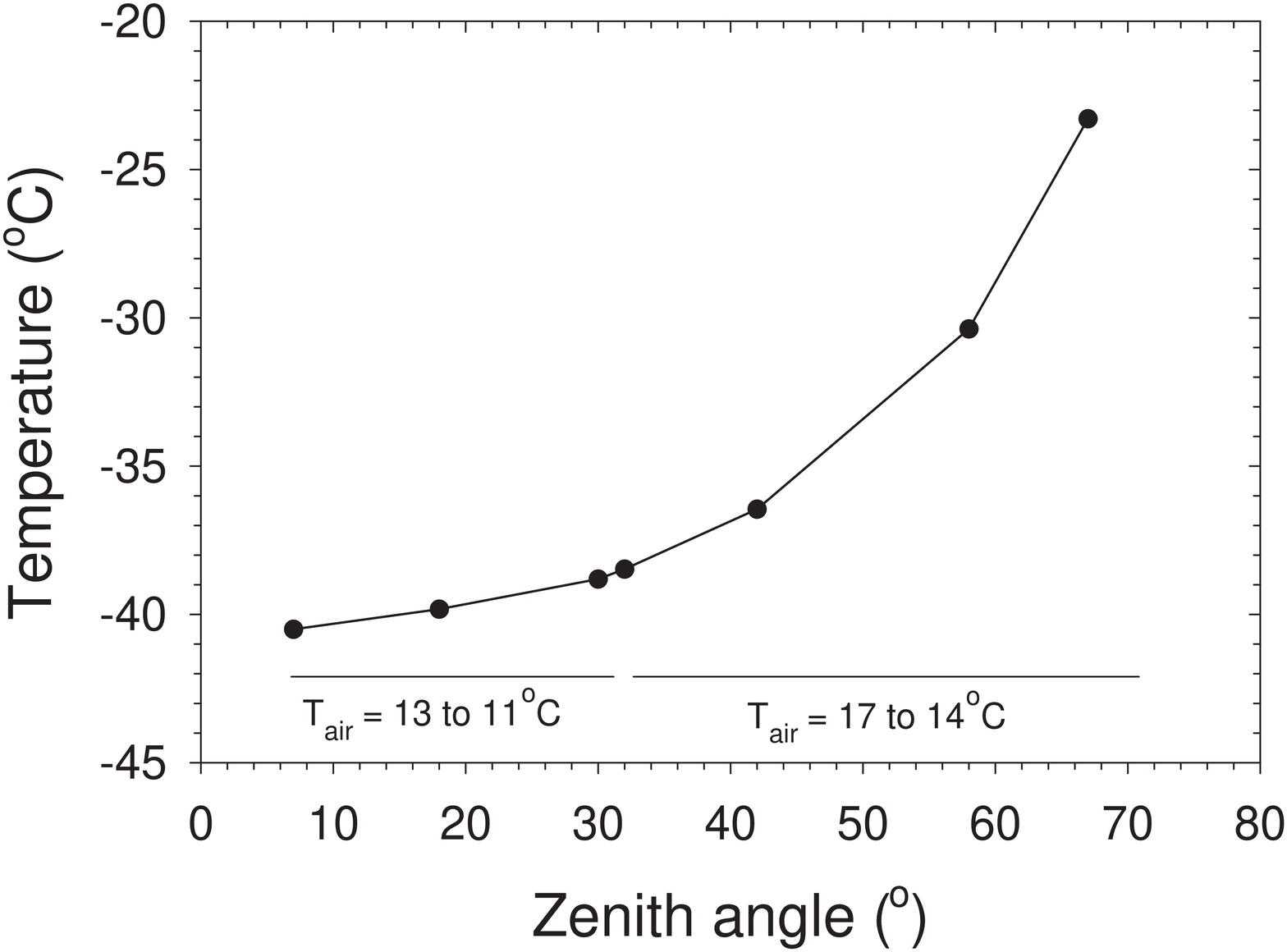,height=8cm}}

\caption{The variation in the clear sky radiometric temperature as the
zenith angle at which the measurement is made is increased from
$0^\circ$ to $67^\circ$. The data were obtained on two nights and there
is a small variation in the air temperature between the observations at
zenith angles $< 30^\circ$ and $> 30^\circ$, as shown in the
figure.}\label{zenith_variation} 

\end{figure}

We show in Figure~\ref{zenith_variation} the measured temperature of a
clear sky using the Narrabri radiometer as a function of zenith angle of
observation in the range $10^\circ - 70^\circ$. The plot is based on
data from observations during two nights when there was a small
difference in air temperature, which is noted. The radiative temperature
of the clear sky increases as the zenith angle is increased and the
measurement is made through an increasing slant thickness of warm lower
atmosphere. The data of the figure also show that the temperature of the
clear sky from night to night is reproducible.

\subsection{Diurnal variation of temperature of a clear sky}

\begin{figure}

\centerline{\psfig{file=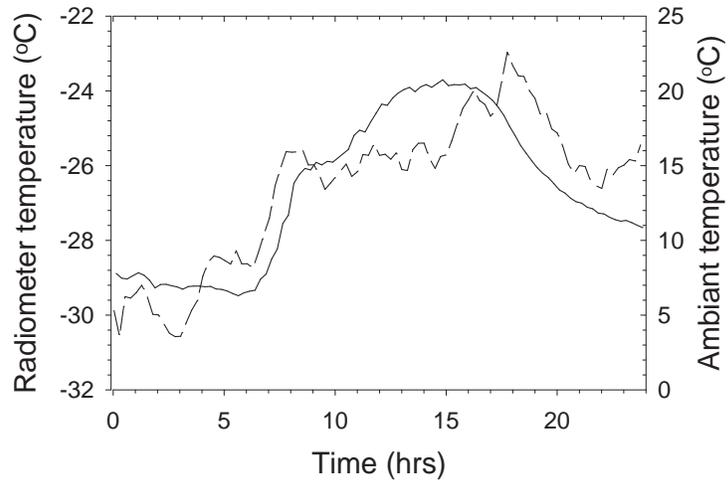,height=8cm}}

\caption{The variation in the clear sky radiometric temperature (solid
line) over a 24 hr period. The profile of the air temperature at screen
height (broken line) is also shown.}\label{daily_variation} 

\end{figure}

The profile of the sky temperature (measured at the zenith near Durham,
England) during a typical 24 hour period during which the sky was clear
is shown in Figure~\ref{daily_variation}. The air temperature at screen
height is also plotted and it is noted that the vertical clear sky
temperature tracks the air temperature at screen height. Such data taken
over a long period are the basis for a simple phenomenological model
linking the observed sky radiometric temperature with the air
temperature and humidity at screen height --- see Section
\ref{sect:idso}.

\subsection{Annual variation of temperature of a clear sky}  

\begin{figure}

\centerline{\psfig{file=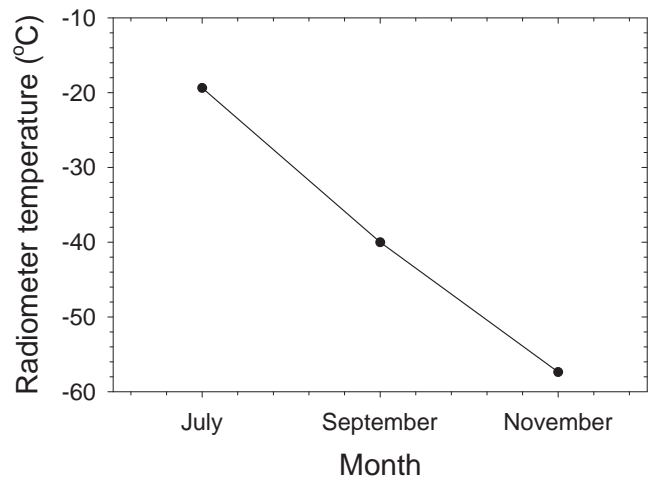,height=8cm}}

\caption{The variation in the clear sky radiometric temperature at the
zenith at 1400 UTC averaged over 4 week periods between from July and
November. The data were recorded near Durham,
England.}\label{yearly_variation} 

\end{figure}

The temperature of a clear sky measured at the zenith at 1400 UTC near
Durham, England averaged over a four week period during July, September
and November is shown in Figure~\ref{yearly_variation}. A systematic
reduction in the sky temperature occurs during the transition from
summer to winter illustrating the annual variation.

\subsection{Temperature of clouds}

\begin{figure}

\centerline{\psfig{file=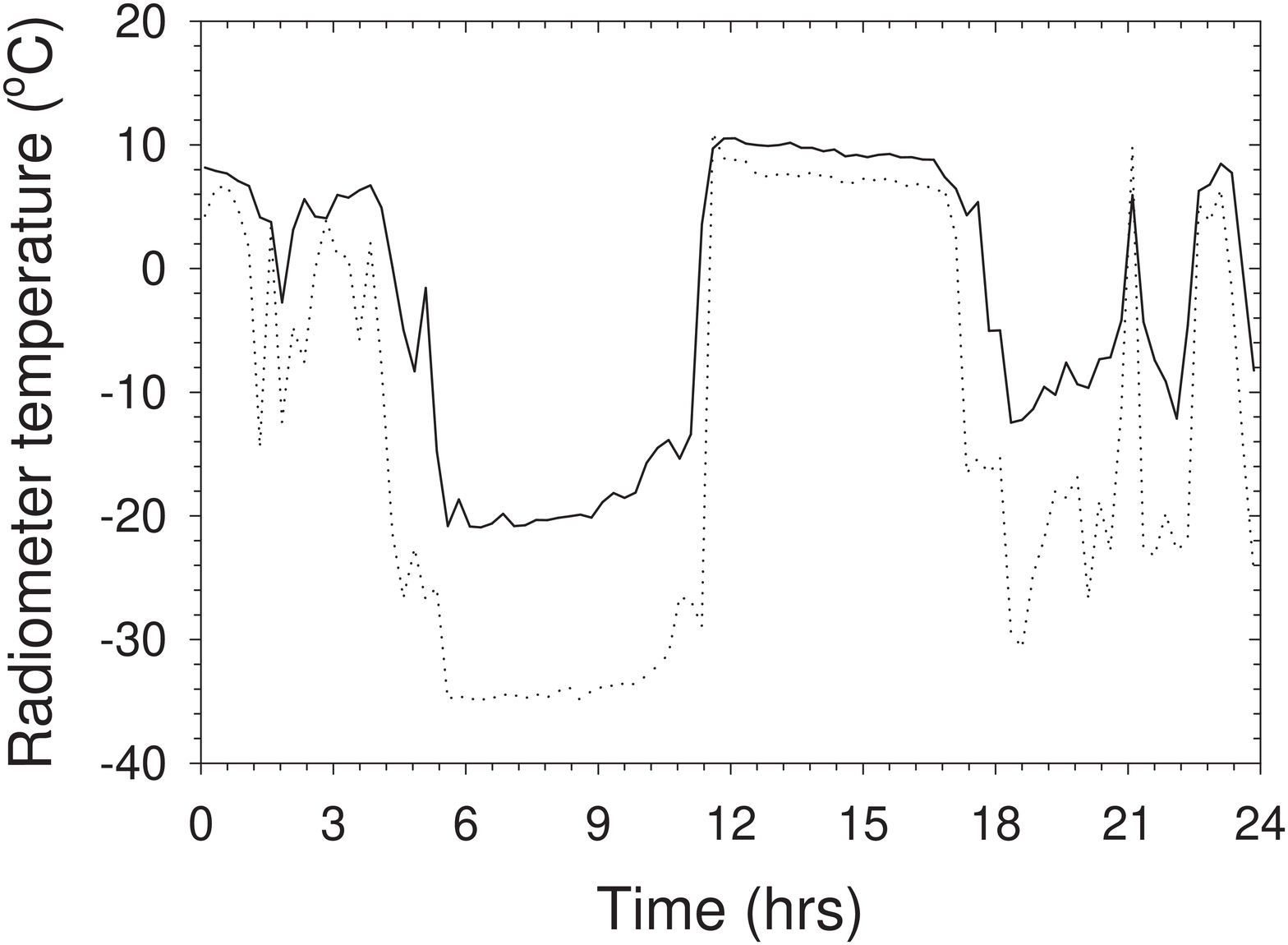,height=8cm}}

\caption{The sky radiative temperatures at the zenith (broken line) and
at $60^\circ$ to the zenith (solid line) in intervals of cloud and clear
sky during a 24 hour period.}\label{clear_and_cloud} 

\end{figure}

Figure~\ref{clear_and_cloud} shows measurements of the sky temperature
at the zenith and $60^\circ$ to the zenith during a 24 hour period when
clear skies were replaced by total overcast clouds with a base at 600 m
at about 1200 UTC. The temperature of the clear sky is seen to be about
$40^\circ$C cooler than than that of the low overcast cloud which is
very similar to the ground level air temperature.

\subsection{Variation of radiometric temperature of clouds with height}

\begin{figure}

\centerline{\psfig{file=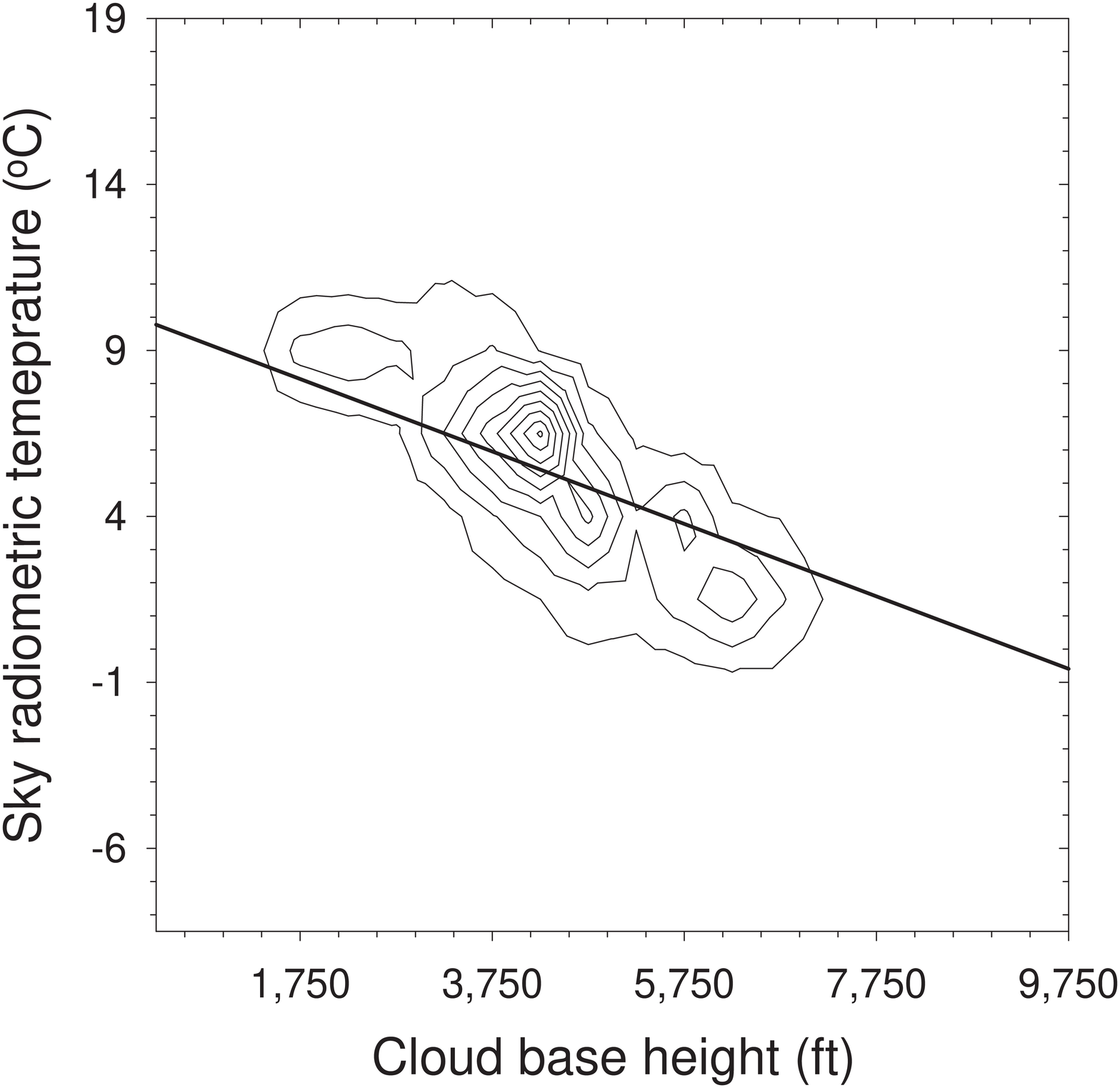,height=8cm}}

\caption{The measured cloud radiative temperature plotted against cloud
base height. The contours specify the frequency of observations of a
given cloud radiative temperature and base height combination. The line
represents the model prediction --- see
Section~\ref{sect:kimball}}\label{scatter_plot} 

\end{figure}

The temperatures of clouds with bases at different heights were measured
at the zenith using a radiometer boresighted with a LIDAR ceilometer ---
see Figure~\ref{scatter_plot}. The radiometer has an aperture of about
$2^\circ$ and on occasion a part of the cold clear sky is viewed when
the cloud base indicated by the narrow angle LIDAR is low. This is the
origin of some spread of the data in Figure~\ref{scatter_plot}. An
increase in cloud base height from 1000 to 7000 ft corresponds to a
decrease in measured cloud temperature from $+10^\circ$C to $0^\circ$C.

\section{A Model for the Temperature of Clear Skies}\label{sect:idso}

\subsection{A simple empirical model}

\begin{figure}

\centerline{\psfig{file=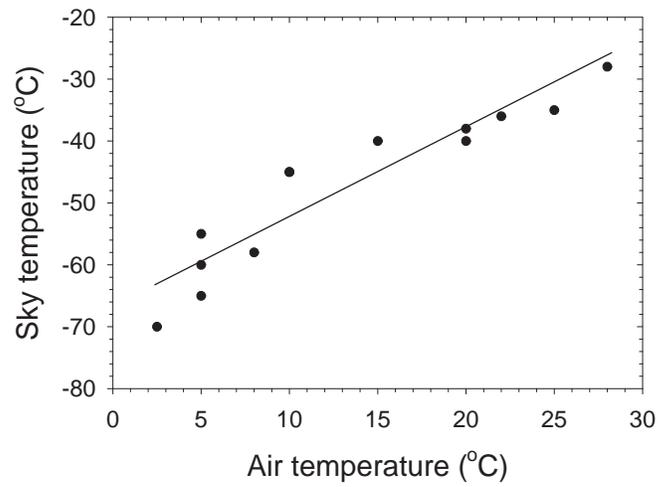,height=8cm}}

\caption{The relationship between the radiative temperature for a clear
sky observed at the zenith and the air temperature at screen
height.}\label{temperature_variation} 

\end{figure}

The data of Figure~\ref{temperature_variation} show the relation between
the radiometric temperature of a clear sky and the air temperature at
screen height for measurements near Durham, England. They suggest that
the major factor governing the clear sky temperature is the temperature
of the air at the observation level. This is representative of the
temperature of the atmosphere through which the measurement is made. 

An approximate representation of the sky radiative temperature is given
by
\begin{equation}
T_{\rm sky} = 1.4 T_0 - 66.2 \label{eq:empirical}
\end{equation}
where $T_{\rm sky}$ is the sky radiative temperature (in $^\circ$C) and
$T_0$ is the air temperature at screen height (in $^\circ$C).

\subsection{The model of Idso (1981)}

A number of models for the emission of long wave radiation from clears
skies have been developed, normally for use with the modelling of
meteorological conditions and in energy-balance models (see e.g.
\citeauthor{Monteith1990}, \citeyear{Monteith1990}). Because of this
principle application, the majority of these models predict the longwave
radiation integrated over a complete hemisphere and not from a small
area of the sky. A recent example of this type of model is that due to
\citeauthor{Prata1996} (\citeyear{Prata1996}). 

\citeauthor{Idso1981} (\citeyear{Idso1981}) described a model which can
be used to predict the radiation from a small area of clear sky. The
model incorporates both the effects of air temperature and vapour
pressure at ground level in the prediction of sky temperatures. 

\citeauthor{Idso1981} showed that the full spectrum radiated by a clear
sky, $R_{\rm a}$, may be represented by
\begin{equation} 
R_{\rm a} = \epsilon_{\rm a} \sigma T_{0}^{4} \label{eq:stefan} 
\end{equation}
where $\epsilon_{\rm a}$ is the full spectrum clear sky emittance, $T_0$
is the screen level air temperature in K and $\sigma$ is the
Stefan-Boltzmann constant.

An equation for $\epsilon_{\rm a}$, appropriate to the 8 -- 14 $\mu$m
wave band at the zenith, was given by \citeauthor{Idso1981} as
\begin{equation} \epsilon_{\rm a}= 0.24 + 2.98 \times 10^{-8} e_{0}^2
\exp(3000/T_0)	\label{eq:predict_ea} \end{equation} where $e_0$ is the
screen level (ground level) vapour pressure and can be calculated using
an equation given by \citeauthor{Wiesner1970} (\citeyear{Wiesner1970}).

\subsection{Validation of the Idso model}

\begin{figure}

\centerline{\psfig{file=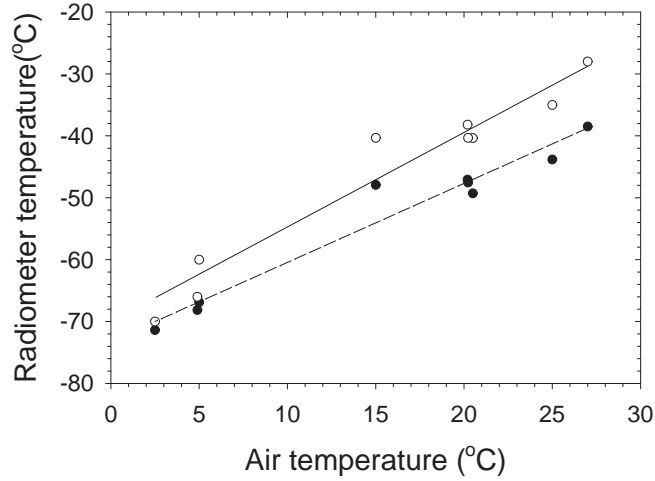,height=8cm}}

\caption{A comparison of the predictions of the Idso model ($\bullet$)
with experimental data ($\circ$) for the variation of radiometer
temperature for clear skies with air temperature at screen height. The
solid line is a linear fit to the experimental data and the broken line
a linear fit to the modelled points.}\label{clear_sky_comparison}

\end{figure}

Data were selected from observations on clear days and the vapour
pressure and air temperatures measured on these days were employed as
input to the model to calculate the radiometer sky temperature. The
predicted sky temperature was compared with the measured values. Figure
\ref{clear_sky_comparison} shows the measured sky temperatures and the
predictions of the model for varying air temperatures at screen height
(the variation in vapour pressures were also allowed for). The predicted
and observed variation of temperature can each be fitted by a straight
line. It can be seen that the predicted and observed temperatures
represented by the straight lines differ by $2^\circ$C at low air
temperatures and $10^\circ$C at high air temperatures. 

This simple model therefore provides a reasonable representation of
temperature of the clear sky. Explanations may exist for the differences
noted. For example, it is possible that the constants of the model
should be adjusted to take account of the difference in climate in the
northeast of England (where our measurements were made) and in Arizona
(where the original data used to validate the Idso model were recorded).
A small adjustment has been made to equation \ref{eq:predict_ea}
specifying the 8 -- 14 $\mu$m clear sky emissivity ($\epsilon_{\rm a}$)
which reduces the values by $\sim 10\%$ and reconciles the predictions
of the model and our measurements. The original formulation by Idso has
been modified to

\begin{equation}
\epsilon_{\rm a} = 0.22 + 2.98 \times 10^{-8} e_{0}^2 \exp(3150/T_0).	\label{eq:mod_idso}
\end{equation} 

\begin{figure}

\centerline{\psfig{file=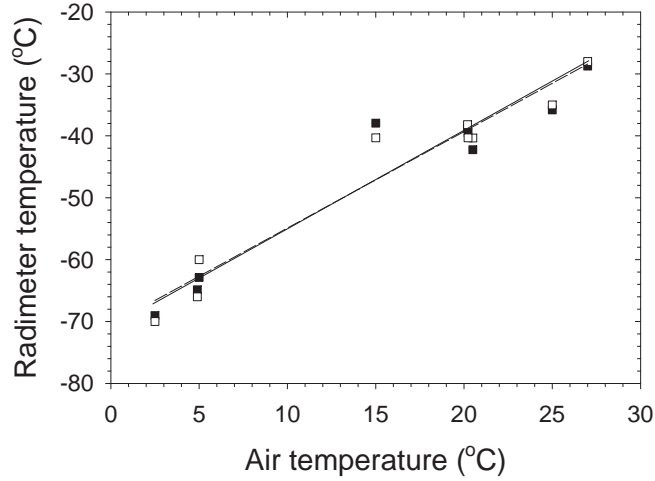,height=8cm}}

\caption{A comparison of the predictions of the modified Idso model
($\blacksquare$) with experimental data ($\square$) for the variation of
radiometer temperature for clear skies with air temperature at screen
height. The predictions (broken line) and the observations (solid line)
are now indistinguishable.}\label{clear_sky_fit}

\end{figure}

This is the form of the equation appropriate for skies observed under
climatic conditions found in northeastern England. Figure
\ref{clear_sky_fit} shows the comparison of the predictions from this
modified model and experimental measurements; the fitted lines are
indistinguishable.

\section{A Model for the Temperature of Clouds}

If clouds are present in the sky, they produce an additional
contribution to the thermal radiation detected at ground level. For a
cloudy sky, therefore, the measured sky radiative temperature is
increased. As is the case for clear skies, much of the discussion in the
literature has been of models for the prediction of the thermal emission
from a complete hemisphere for use in meteorological models (e.g.
\citeauthor{Alados1995}, \citeyear{Alados1995}; \citeauthor{Malek1997},
\citeyear{Malek1997}). However \citeauthor{Kimball1982}
(\citeyear{Kimball1982}) have developed a model that can be used to
predict the cloud temperature at a specified zenith angle.

\subsection{The model of Kimball et al. (1982)}

This model for the emission from a cloudy sky is based on the clear sky
model of \citeauthor{Idso1981} (\citeyear{Idso1981}) but with an
additional term to include radiation effects of the cloud. In this model
the contribution to the thermal radiation from clouds, $R_{\rm c}$, is
\begin{equation} R_{\rm c} = \tau A \epsilon_{\rm c} f \sigma T_{\rm c}^4 \label{eq:kimball} 
\end{equation} 
where $\epsilon_{\rm c}$ is the cloud emittance (a value dependant on
cloud type in the range 0.5 to 1.0), $f$ is the fraction of black body
radiation emitted in the 8 -- 14 $\mu$m band, (taken as 1.0 in this
work), $A$ is the fraction of sky covered with cloud, $\sigma$ is the
Stephan-Boltzmann constant and $\tau$ is the transmittance of the
atmosphere in the 8 -- 14 $\mu$m window where
\begin{equation} 
\tau = 1 - \epsilon_{\rm z}	\label{eq:define_t}
\end{equation} 
and $\epsilon_{\rm z}$ is the 8 -- 14 $\mu$m sky emittance
\cite{Idso1981} --- see equation \ref{eq:predict_ea}. 

$T_{\rm c}$ is the cloud temperature, which is calculated assuming the
standard temperature lapse rate (see e.g. \citeauthor{Lutgens1998},
\citeyear{Lutgens1998}). According to this the equation for the
temperature of the cloud is
\begin{equation}
T_{\rm c} = T_0 - 0.0065 Z \label{eq:predict_cloud_temp} 
\end{equation}
where $T_0$ is the air temperature at screen
height in K and $Z$ is the height of the cloud in meters. 

The total predicted sky thermal radiation, $R_{\rm T}$, received at the
earth's surface from the cloudy sky is then
\begin{equation}
R_{\rm T} = R_{\rm a} + R_{\rm c}
\end{equation} 
where $R_{\rm a}$ is the clear sky radiation --- see equation
\ref{eq:stefan}.

\subsection{Validation of the model} \label{sect:kimball}

\begin{figure}

\centerline{\psfig{file=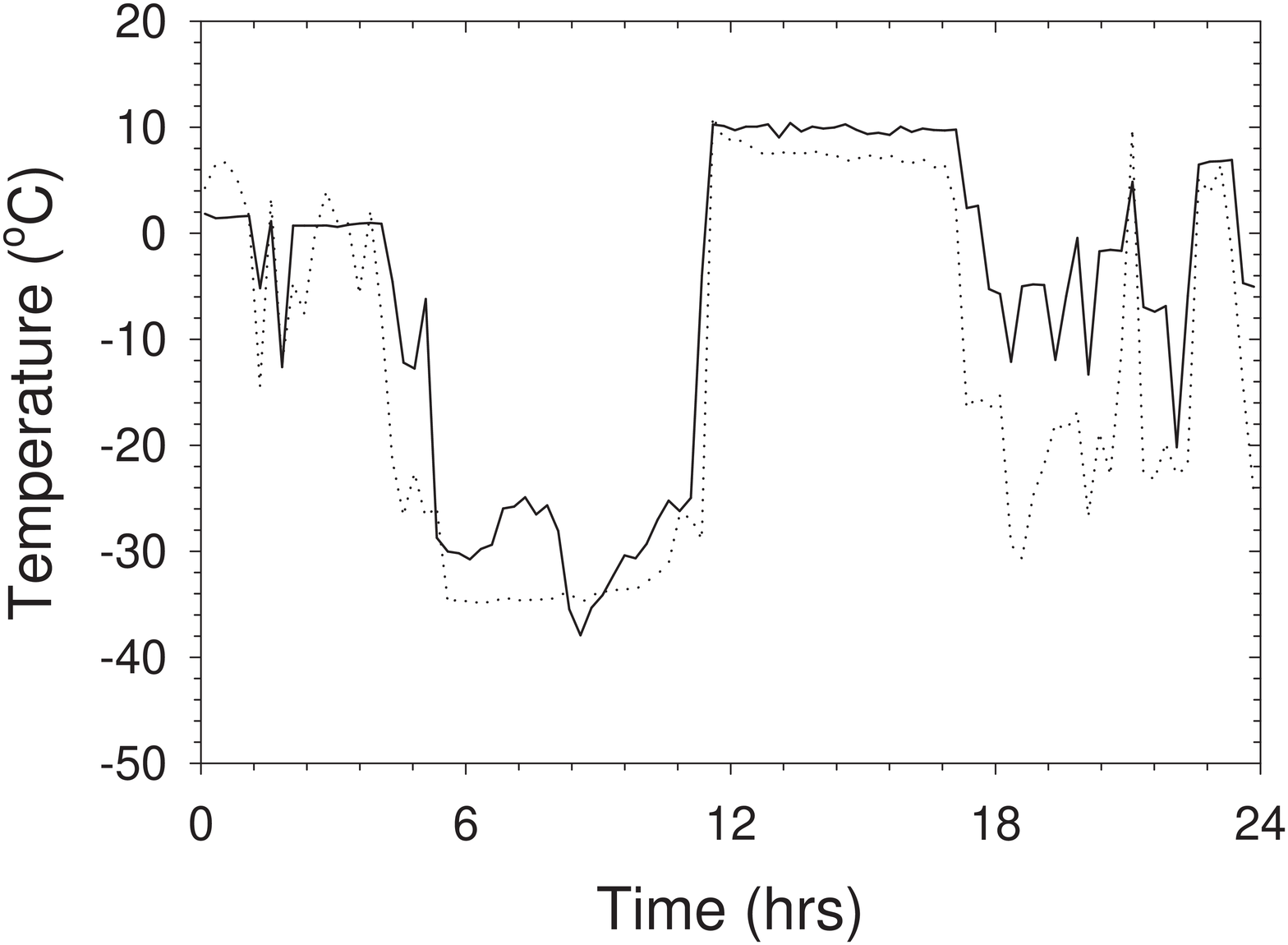,height=8cm}}

\caption{A comparison of the predictions of the Kimball et al. model
(solid line) with observed cloud/sky temperature (broken line) during a
changeable day.}\label{cloudy_sky_fit}

\end{figure}

The cloud model developed by \citeauthor{Kimball1982}
(\citeyear{Kimball1982}) was used to predict radiometer temperatures for
cloudy skies which may be compared with observations. Figure
\ref{cloudy_sky_fit} shows a comparison between the predictions of the
model and observations on a day which included changes in cloud cover.
The data were taken in August in northeast England. The best fit of the
model predictions and observations (which is shown) was when the cloud
emissivity was taken to be 0.8 (a reasonable mid-range value).

Comparison of the predictions of the \citeauthor{Kimball1982} model with
the measurements of temperature for clouds at different heights are
shown in Figure \ref{scatter_plot}. There is good agreement between the
solid line representing the predictions and the contours repesenting the
data when a value of 0.8 is used for the cloud emissivity.

\section{Summary}

The use of a FIR radiometer sensitive to the 8 -- 14 $\mu$m waveband
mounted on a ground based gamma ray telescope provides a simple and
inexpensive method of reliably monitoring the atmospheric clarity in the
telescope's field of view. The response of the radiometer is prompt and
is independent of the performance of the telescope itself.

The predictions of simple semi-empirical models of the longwave emission
from clear skies \cite{Idso1981} and cloudy skies \cite{Kimball1982}
have been compared with our measurements. They provide theoretical
underpinning for the observed variation of radiative temperature of the
sky and clouds with the air temperature and relative humidity at screen
height.

\acknowledgements

We are grateful to the UK Particle Physics and Astronomy Research
Council for support of the gamma ray project and Muir Matheson Limited
for the loan of the LIDAR system. Many members of the Gamma Ray
Astronomy Group in the Department of Physics contributed to the
collection of data in Australia. Observations made in Northeast England
were part of a program conducted by the Industrial Research
Laboratories, University of Durham on behalf of the Science and
Technology Division, Defence Clothing and Textiles Agency. These data
are reproduced with the permission of the Director, Science and
Technology Division, Defence Clothing and Textiles Agency.


\begin{thebibliography}{}

\bibitem[\protect\citeauthoryear{Abu-Zayyad et
al.}{1997}]{Abuzayyad1997}Abu-Zayyad, T., Al-Seady, M., Belov, K., Bird,
D. J., Boyer, J., Chen, G., Clay, R. W., Dai, H. Y., Dawson, B. R., Ho,
Y., Huang, A., Jui, C. C. H., Kidd, M., Kieda, D. B., Knapp, B., Lee,
W., Loh, E. C., Mannel, E. J., Matthews, J. N., O'Halloran, T., Salman,
A., Simpson, K. M., Smith, J. D., Sokolsky, P., Sommers, P., Taylor, S.,
Thomas, S.B., Wiencke, L. R., and Wilkinson, C. R.: 1997 {\em Proc. 25th
Int. Cosmic Ray Conf., Durban}, {\bf 5}, 345--348.

\bibitem[\protect\citeauthoryear{Alados-Arboledas et
al.}{1995}]{Alados1995}Alados-Arboledas, L., Vida, J. and Olmo, F. J.:
1995, {\em Int. J. Climatology}, {\bf15}, 107--116.

\bibitem[\protect\citeauthoryear{Armstrong et
al.}{1998}]{Armstrong1998}Armstrong, P., Chadwick, P. M., Cottle, P. J.,
Dickinson, J. E., Dickinson, M. R., Dipper, N. A., Hilton, S. E., Hogg,
W., Holder, J., Kendall, T. R., McComb, T. J. L., Moore, C. M., Orford,
K. J., Rayner, S. M., Roberts, I. D., Roberts, M. D., Robertshaw, M.,
Shaw, S. E., Tindale, K., Tummey, S. P. and Turver, K. E.: 1998, {\em
Exp. Astron.}, in the press.

\bibitem[\protect\citeauthoryear{Ashley and
Jurcevic}{1991}]{Ashley1991}{Ashley, M. C. B. and Jurcevic, J. S.: 1991,
{\em Proc. Astron. Soc. Australia}, {\bf 9}, 334--335.

\bibitem[\protect\citeauthoryear{Bird et al.}{1997}]{Bird1997}Bird, D.
J., Clay, R. W., Dawson, B. R., Gregory, A. G., Smith, A. G. K.,
Johnston, M. and Wild, N. R.: 1997, {\em Proc. 25th Int. Cosmic Ray
Conf., Durban}, {\bf 5}, 353--356.

\bibitem[\protect\citeauthoryear{Fegan}{1997}]{Fegan1997}Fegan, D. J.:
1997, {\em J. Phys. G.: Nucl. Part. Phys.}, {\bf 23}, 1013--1060.

\bibitem[\protect\citeauthoryear{Idso}{1981}]{Idso1981}Idso, S. B.:
1981, {\em Water Resour. Res.}, {\bf17}, 295--304.

\bibitem[\protect\citeauthoryear{Kimball et
al.}{1982}]{Kimball1982}Kimball, B. A., Idso, S. B. and Aase, J. K.:
1982, {\em Water Resour. Res.}, {\bf 18}, 931--936.

\bibitem[\protect\citeauthoryear{Lutgens and
Tarbuck}{1998}]{Lutgens1998}Lutgens, F. K. and Tarbuck, E. J.: 1998 {\em
The Atmosphere}, (Prentice Hall, New York), 15.

\bibitem[\protect\citeauthoryear{Malek}{1997}]{Malek1997}Malek, E.:
1997, {\em Atmos. Res.}, {\bf 45}, 41--54.

\bibitem[\protect\citeauthoryear{Monteith and
Unsworth}{1990}]{Monteith1990}Monteith, J. L. and Unsworth, M. H.: 1990,
{\em Principles of Environmental Physics}, 2nd. ed., (Edward Arnold,
London).

\bibitem[\protect\citeauthoryear{Prata}{1996}]{Prata1996}Prata, A. J.:
1996, {\em Q. J. R. Meteorol. Soc.}, {\bf 122}, 1127--1151.

\bibitem[\protect\citeauthoryear{Wiesner}{1970}]{Wiesner1970}Wiesner, C.
J.: 1970, {\em Hydrometeorology}, (Chapman and Hall, London), 43.

}
\end{thebibliography}
\end{document}